\documentclass[aps,prb,superscriptaddress,showpacs,floatfix,twocolumn]{revtex4-1}
\usepackage{times}
\usepackage{graphicx,graphics,color}% Include figure files
\usepackage{amsmath}
\usepackage{amssymb}
\usepackage{ulem}
\usepackage{color}

\begin{document}

\title{An efficient clustering algorithm from the measure of local Gaussian distribution}

\author{Yuan-Yen Tai}

\date{\today}

\begin{abstract}
In this paper, I will introduce a fast and novel clustering algorithm based on Gaussian distribution and it can guarantee the separation of each cluster centroid as a given parameter, $d_s$.
The worst run time complexity of this algorithm is approximately $\sim$O$(T\times N \times \log(N))$ where $T$ is the iteration steps and $N$ is the number of features.
\end{abstract}

\maketitle

\section{Introduction}
Clustering algorithms have many applications~\cite{algbook,algsurvey} to the unsupervised and semisupervised learning, which they help classify data sets into clusters within few given parameters.
Potentially, it can be applied to many different domains such as image analysis, object tracking, data compression, physical/chemical structure optimization problems, to name a few.
Some cluster algorithms (e.g. K-means~\cite{kmean} and Gaussian-Mixture~\cite{gmmbook}) require the knowledge of number of clusters as input parameters, and some others (e.g. Dbscan~\cite{dbscan}, Hdbscan~\cite{hdbscan}) require a distance cut-off threshold to separate clusters.
Each clustering algorithm has its own pros and cons and its own domain for different applications.
My goal in this paper is to invent a new clustering algorithm which has the following properties:
\begin{itemize}
\item Require each cluster to be a Gaussian-like distribution.
\item Free from the cluster number parameter.
\item Able to deal with large number of data set efficiently.
\end{itemize}
Over all, this new clustering algorithm is able to automatically search for cluster numbers with a given separation threshold in a fast and robust way, which was difficult for {\bf K-means} and {\bf Gaussian-Mixture}.
While, if all clusters are gaussian-like, it is also more efficient than {\bf Hdbscan} in finding major clusters automatically.
The reason for such efficient gain is based on the design of the new algorithm which is using local data information instead of compare each data pair in a global fashion.
Nevertheless, this algorithm can be regarded as an efficient improvement between Gaussian-Mixture model and K-means, and an improvement of Hdbscan can be found in here~\cite{hdbscan2}.

\section{The Algorithm}
This section will walk through the concepts of this algorithm, and I will analyze more in-depth details in the next section.
Over all, the entire algorithm is going through the following analysis pipeline:
\begin{itemize}
\item[a.] Indexing all data points by the R-tree structure.
\item[b.] Seeding centroids across K-dimensional features.
\item[c.] Converge centroids and delete excessive ones to find the definitive cluster centroid.
\item[d.] Iteratively calculate co-variant matrix, $\Sigma_c$, from weighted local data points.
\item[e.] Re-assign data points to each cluster id via the Gaussian distribution, P$(x|\vec\mu_c)$.\\
\end{itemize}
Where steps b. and c. are designed for finding the definitive centroid of each cluster.
Once the definitive cluster centroid has been found, the next step is to calculate co-variance matrix of the Gaussian distribution via a mean-field iterative calculation in step D.
Finally, re-assign all the data points to the correspond cluster id in step e.
The following paragraphs describe more details of the entire algorithm.\\

\paragraph{\bf Indexing all data points by the R-tree structure.}
The very first step is to construct the K-dimensional R-tree with all the data points, $\vec x_i$, which generally gives average O$(\log N)$ complexity of multi-dimensional range search.
Here, the index $i$ indicated the $i$-$th$ data point.
During the construction of R-tree, the minimum ($x_{\min}^a$) and maximum ($x_{\max}^a$) value of feature, $a$, can also be found without loosing too much calculation resource.\\

\begin{figure}
\includegraphics[scale=0.16,angle=0]{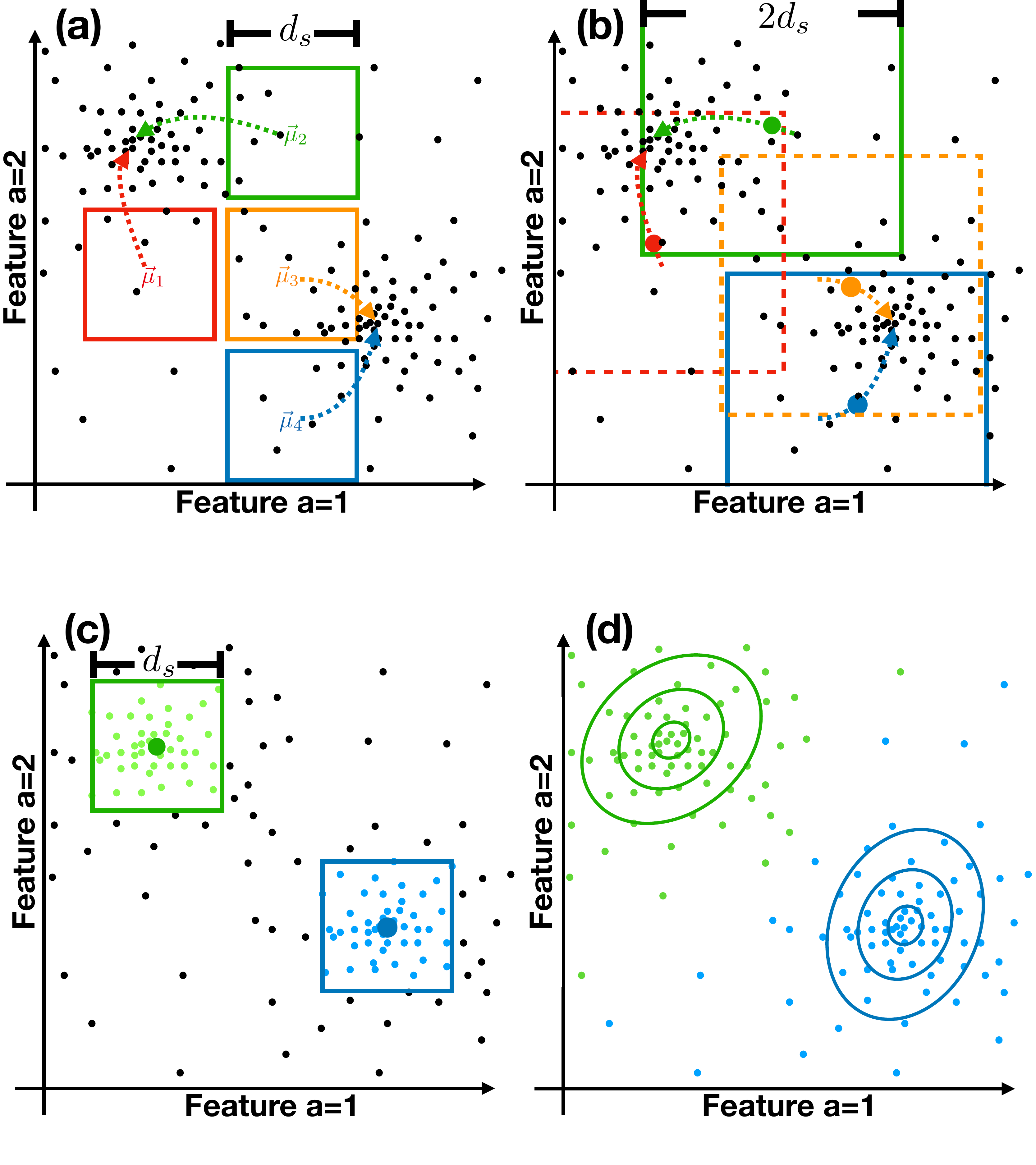}
\caption {
The illustration of the iterative convergence of four centroid seed into two clusters in a K=2 (2 dimensional) data set.
In (a), the cubes (squares) indicated the local searching $d_s^{K=2}$ volume of each centroid.
In (b), the cubes (squares) indicated the $d_c \equiv 2d_s$ collision box of each centroid.
While, the position of centroids are moved a little bit from (b) to (a) to illustrate the iterations of step c. in the algorithm.
After finding the definitive centroids, the local data points has been collected as shown in (c), and used for the calculation of co-variant matrixs, $\Sigma_c$, to fit the distribution, $P(\vec x, \vec\mu_c)$, as shown in (d).
}\label{fig01}
\end{figure}

\paragraph{\bf Seeding centroids across K-dimensional features.}
Now, the centroids, $\vec\mu^{t=0}_c$, are seeded with a given separation distance, $d_s$, across all features within $[x_{\min}^a, x_{\max}^a]$, where $t=0$ means the zeroth iteration of the centroid.
Fig.~\ref{fig01}.a shows the seeding of centroids, where $\vec\mu_1$ is the first centroid inside the red cube(square) and so on so forth.

During the seeding process, the algorithm also search for the data points in the K-dimensional cubic of volume, $d_s^K$, around each centroid $\mu_c$, and thus find out the temporary local data sets, $\vec x_{i \in c}$, to the correspond centroid.
Note that, each $\vec\mu_c$ possess a local count of data points, $N_c$, inside each $d_s^K$ volume.
By this nature, several centroids can be excluded if they possess too few counts of data point by a given counting threshold, $N_c < L$.\\

\paragraph{\bf Converge centroids and delete excessive ones to find the definitive cluster centers.}
The previous step can be regarded as the zeroth iteration of the converge process.
In order to find the best cluster centers, centroid need to be updated, $\vec\mu_c$, by calculate the means of $\vec x_{i\in c}$,
\begin{equation}
\vec\mu^{t+1}_c = \frac{1}{N_c}\sum_{i\in c} \vec x_{i}.
\end{equation}
After words, the iterations of each centroid can be stopped by the criterion,
\begin{equation}
|\vec\mu_c^{t+1} - \vec\mu_c^{t}| < \epsilon,
\end{equation}
where $|\cdots|$ denotes the Euclidean distance under K-dimension and $\epsilon$ is the convergence threshold.
In Fig.~\ref{fig01}.a, the dotted lines show the iteration paths of each centroid, $\vec\mu_c$.

During the iteration of centroids, two or more centroid seeds may ends up in the same cluster center.
According to this factor, the iteration process can be further speed up by delete the centroid $\mu_c$ with lower data point counting, $N_c$, within a collision detection.
The collision detection is defined in a K-dimensional cubic box that spanned by the collision distance, $d_c \equiv 2d_s$.
Fig.~\ref{fig01}.b shows $\mu_1$ (red) and $\mu_3$ (orange) to be deleted as they possess less count ($N_c$) than $\mu_2$ (green) and $\mu_4$ (blue).

Note that, it is possible to set a different value of $d_c$ instead of using $d_c \equiv 2d_s$.
However, it would be efficient and accurate enough by setting $d_c \equiv 2d_s$.
In principal, there could be several different ways to implement the centroid deletion algorithm.
However, I will not dive into too much details in here.\\

\begin{figure}
\includegraphics[scale=0.33,angle=0]{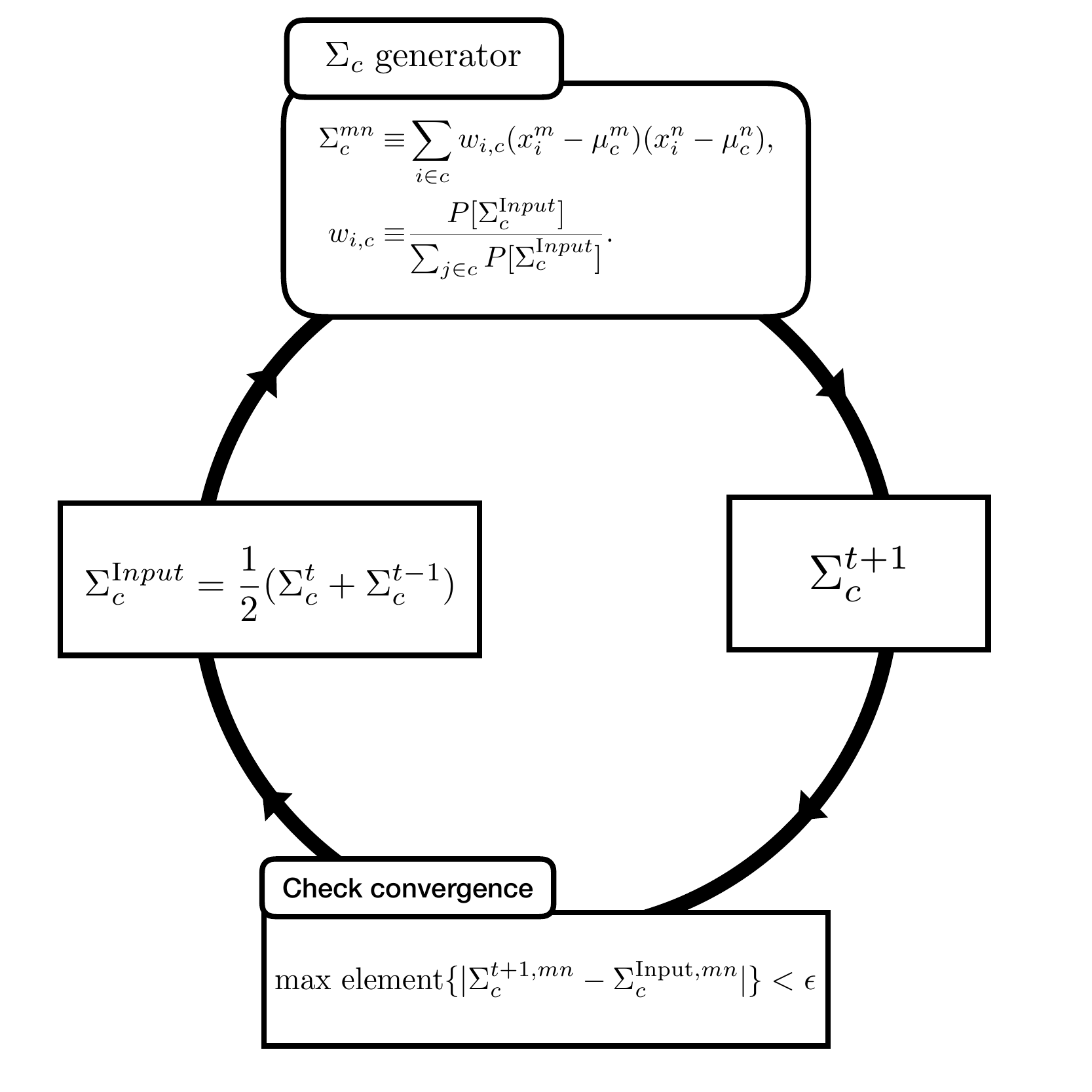}
\caption {
The self-consistent loop calculation for $\Sigma_c$ for the process of ii. and iii. in step d. of the algorithm.
}\label{fig02}
\end{figure}

\paragraph{\bf Iteratively calculate co-variant matrix, $\Sigma_c$, from weighted local data points.}
After finding all the definitive cluster centroid, $\mu_c$, the co-variant matrix $\Sigma_c$ of each cluster are ready to be calculated via the local data points, $x_{i\in c}$.
Firstly, the Gaussian distribution is generally written,
\begin{equation}
P(\vec x|\mu_c) = \frac{1}{\sqrt{2\pi |\Sigma_c|}} \times e^{-(\vec x - \vec\mu_c)^T \Sigma_c^{-1} (\vec x - \vec\mu_c)},
\label{gauss}
\end{equation}
where, $\vec x$ and $\vec\mu_c$ are column vectors and the $T$ operation is the transpose of them.
The following formula is the usual way to calculate the co-variant matrix from a given data set,
\begin{equation}
\label{eq0}
\Sigma^{mn}_c = \frac{1}{N_c} \sum_{i\in c} (x^m_i-\mu^m_c)(x^n_i-\mu^n_c).
\end{equation}
While, since $x_{i\in c}$ is not a complete data set from each cluster centroid $\mu_c$, it might induce inaccuracies.
Therefore, Eq.~\ref{eq0} can be modified into the following equations,
\begin{equation}
\begin{split}
\label{eqn}
\Sigma^{mn}_c \equiv& \sum_{i\in c} \frac{w_{i,c}}{w_{i,c}+\eta} (x^m_i-\mu^m_c)(x^n_i-\mu^n_c),\\
w_{i,c} \equiv& \frac{P(\vec x_i| \vec \mu_c)}{\sum_{j\in c} P(\vec x_j| \vec \mu_c)}.
\end{split}
\end{equation}
Here $\eta$ is a small number that can be choosed by experience between 0 and 1.
It means, some data points of $\vec x_{i\in c}$ get more important if they are closer to the cluster centroid, $\vec\mu_c$, according to the weight factor, $w_{i,c}$, where if the likelihood of a data point is close or smaller to $\eta$, the weighting will be much surpressed by the effect of,
\begin{equation}
\frac{w_{i,c}}{w_{i,c}+\eta}
\end{equation}
Overall, Eq.~\ref{eqn} meant to keep good quality of the co-variant matrix, $\Sigma_c$, that even it is only calculated via the local data set, $\vec x_{i\in c}$.
However, Eq.~\ref{eqn} also indicated a functional form, $P(\vec x_i| \vec \mu_c) \equiv P_i\big[\Sigma_c\big]$, and the following variational condition can be carried out to yield the optimization of $P[\Sigma_c]$,
\begin{equation}
\label{eqvari}
\delta \int d\vec x\, P[\Sigma_c] = 0.
\end{equation}
A mathematical rigorous solution for Eq.~\ref{eqvari} might be hard to get.
While, Eq.~\ref{eqn} and ~\ref{eqvari} can be approximately solved by mean-field method, iteratively,
\begin{itemize}
\item[i.] Calculate $\Sigma_c^0$ from Eq.~\ref{eq0}, and set $\Sigma_c^1 = \Sigma_c^0$,

\item[ii.] Calculate $\Sigma_c^{t+1}$ from Eq.~\ref{eqn} with the input of the co-variant matrix for $P[\Sigma_c^{\text Input}]$, where $\Sigma_c^{\text Input} = \frac{1}{2}(\Sigma_c^{t}+\Sigma_c^{t-1})$,

\item[iii.] Iterate process ii. until,
\begin{equation}
\text{max element}\{|\Sigma_c^{t+1,mn} - \Sigma_c^{\text{Input}, mn}|\} < \epsilon,
\end{equation}
\end{itemize}
where $\epsilon$ is the convergence threshold.
In above, the form of the second step is to ensure a smooth iteration process for the co-variant matrix.
The entire process of the self-consistence loop calculation is shown in Fig.~\ref{fig02}.\\

\paragraph{\bf Re-assign data points to each cluster center via the Gaussian distribution, P$(\vec x|\vec\mu_c)$.}
Finally, after all of the co-variant matrix, $\Sigma_c$, for each centroid are calculated, the assignment of each data point, $\vec x_i$, become very easy.
%We re-assign each data point, $\vec x_i$, to the largest P-value of the centroid,
\begin{equation}
\begin{split}
\vec x_i& \in \vec\mu_{\text{max}} \text{, where }\\
\vec\mu_{\text{max}}& \gets \text{max P-value of } \{P(\vec x_i|\vec\mu_0), ..., P(\vec x_i|\vec\mu_n)\}.
\end{split}
\end{equation}
The entire process of the clustering algorithm ends here.\\
%However, due to the nature of the Gaussian distribution, $P(\vec x| \vec\mu_c)$, some post-process can be easily defined to exclude some data points and ensure the LoD (limit of detection).

\section{Discussions to the Algorithm}
This section is divided into three sub-sections, where I will discuss: {\bf A.} how to set proper parameters for the clustering algorithm, {\bf B.} analyze the run time complexities of the analysis pipeline, and {\bf C.} establish some post clustering process to ensure the read out data quality.

\begin{figure}
\includegraphics[scale=0.16,angle=0]{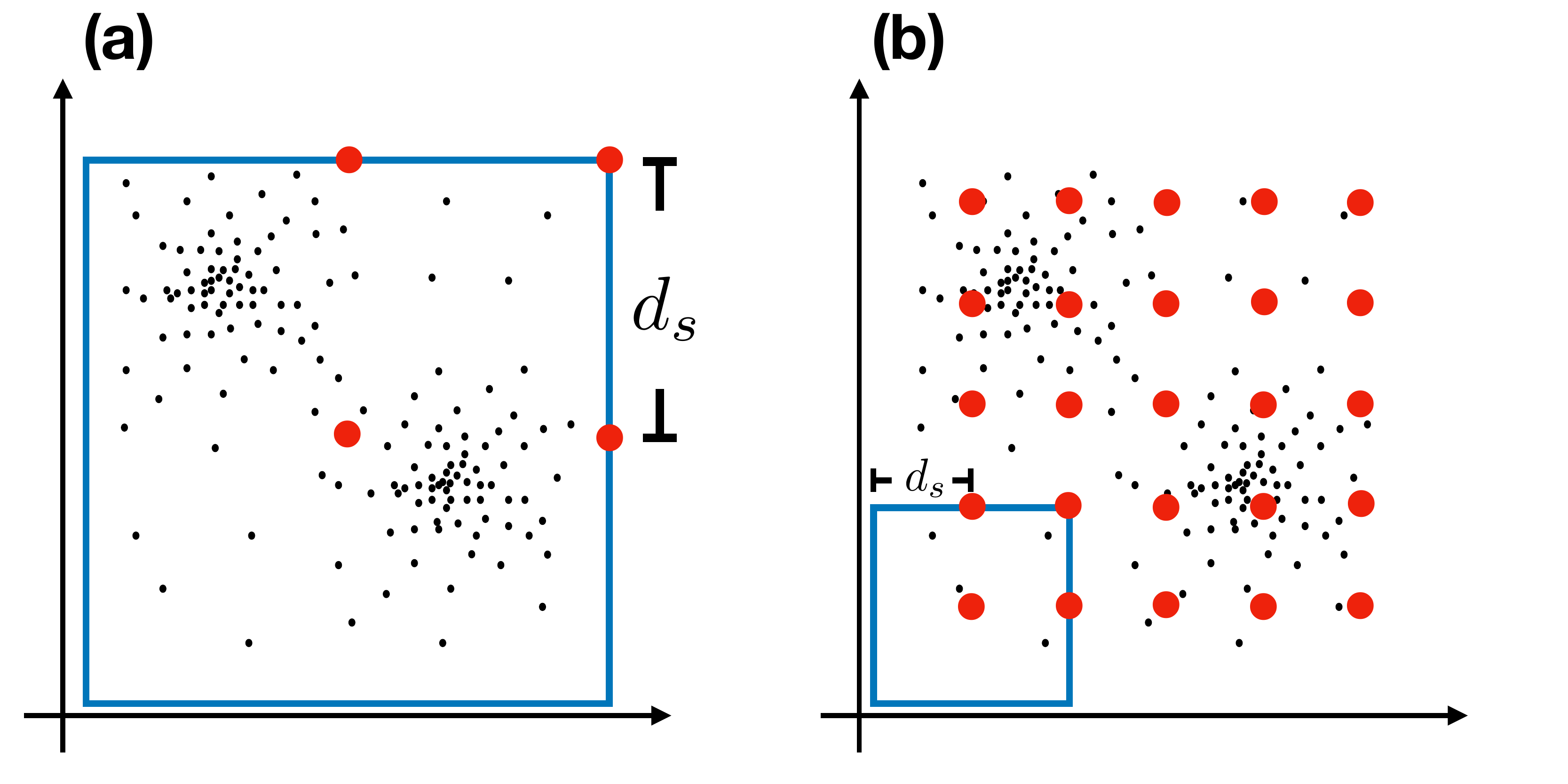}
\caption {
Two different settings for the centroid separation distance, (a) larger separation, (b) smaller separation.
The red circle presenting the centroid seeds, and the blue cube (square) indicated the collision box.
}\label{fig03}
\end{figure}

\subsection{Parameter settings}
Three parameters was mentioned in ther previous section that describe the algorithm:
\begin{itemize}
\item The centroid separation distance, $d_s$.
\item The local data counting threshold, $L$.
\item The convergence threshold, $\epsilon$.
\end{itemize}
The convergence threshold, $\epsilon \sim 0.01$, is a small number, and it is not sensitive in general.
The local data counting threshold, $L$, is an empirical parameter which depends on the amount of data points and the separation distance, $d_s$.
$L$ only served the purpose to boost the initialization and iteration of the algorithm during in steps b. and c., and the results will not be changed that even $L=0$.
A good choice of $L$ can boost the calculation speed as well as eliminate some small clusters in the beginning.
While the centroid separation distance $d_s$ is an important parameter, if it is wrongly set, the finding of clusters could be changed.
If a larger value of $d_s$ is given, illustrated in Fig.~\ref{fig03} (a), the calculation could be faster due to fewer count of centroid seeds in the initial stage.
But it is possible that only a single cluster centroid to be survived due to a large collision distance, $d_c\equiv 2d_s$.
On the other hand, if a smaller value of $d_s$ is given, illustrated in Fig.~\ref{fig03} (b), the calculation will be slower a bit due to more count of centroid seeds, yet two cluster can be found.
However, if an extremely small value of $d_s$ has been set, it is possible to found each cluster that only possess a single data point, which is undesirable.
A simple rule of thub can be applied to make a good choice of $d_s$, where $d_s$ can be set close to but smaller than one-half of the smallest distance from the actual cluster centers.\\

\subsection{Run time complexities}
It is complicate to establish a precise analysis for the run time complexity due to three inter-related variables: (a) number of {\bf features}, (b) number of {\bf data points} and (c) number of {\bf clusters}.
Therefore, I will focus on the special situation for ``small number of feature" with ``few clusters".
Here I list the worst run time complexity of each step for this situation,
\begin{itemize}
\item[a.] Indexing all data points by the R-tree structure:\\
$\sim O(N \times \log (N))$,
\item[b.] Seeding centroids across K-dimensional features:\\
$= \sum_{c=0}^M N_c \times \log(N) \sim O(N \times \log(N))$ ,
\item[c.] Converge centroids and delete excessive ones to find the definitive cluster centroid:\\
$\sim O(T_{\mu_c} \times N_c \times \log(N))$,
\item[d.] Iteratively calculate co-variant matrix, $\Sigma_c$, from weighted local data points:\\
$\sim O(T_{\Sigma_c} \times N_c \times K^3)$,
\item[e.] Re-assign data points to each cluster id via the Gaussian distribution, P$(x|\vec\mu_c)$:\\
$\sim O(N_c)$,
\end{itemize}
where $N$ is the total number of data points, $M$ is the number of centroid seeds, $T_{\mu_c}$ is the centroid iteration steps, $N_c$ is the number of local data points that belong to $\mu_c$, $T_{\Sigma_c}$ is the iteration steps in d., and K is the dimension (feature) of the data.

Note that, a matrix-inversion operation is required for Eq.~\ref{gauss}, and it takes $O(K^3)$ run time complexity.
If there are only few features (small number of K) to be considered, this operation could be regarded as constant run time complexity.
Finally, it is easy to know the bottleneck of the entire algorithm is in either c. or d. which depends on the types of data sets.
Therefore, the upper limit run time complexity could be roughly estimated, $\sim O(T_\text{max}\times N \times \log(N))$, where the maximum of iterations, $T_{\text max}$, is less then 20 in most of the cases.

\subsection{Post clustering process}
\begin{figure}
\includegraphics[scale=0.22,angle=0]{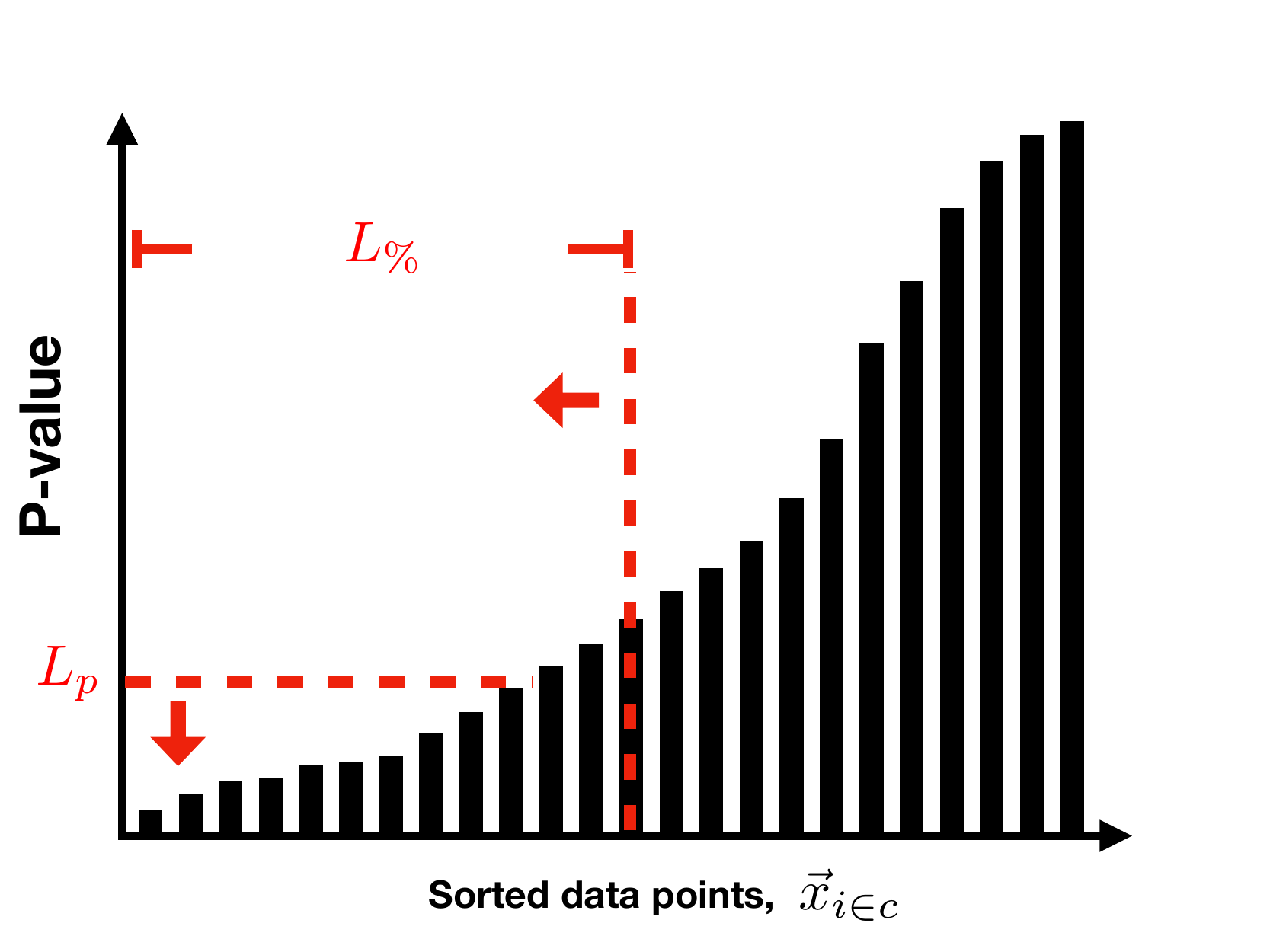}
\caption {
Illustration of the definition of $L_p$ and $L_\%$ for a given cluster, $\vec x_{i\in c}$.
The red arrow indicated the portion of data to be dropped.
}\label{fig03}
\end{figure}

Clustering algorithms serve in many different purpose of usage.
For example, in a cloth store, the salesman can apply some clustering algorithms to suggest a customer to buy which size of cloth based on their hight and weight.
In this situation, almost all data points (customers) should be considered, and thus to be assigned to the correspond cluster centroids (the size labels).
However, it is not a good idea to include all data points in some other situations, where some falsely classified data points need to be avoid based on experimental facts.
In this situation, it is more prefferable to take data points which are closer to the cluster centroid.
If we are dealing with the second scenario, the benefit of the Gaussian distribution become obvious.
Due to the nature of Gaussian distribution, the P-value can be easily traced with a good model fit, and Gaussian-Mixture model (GM) was invented to serve this purpose.
However, GM is not able to deal with large number of data points due to its time complexity, $\sim O(T\times N^2)$, and this is one of the reasons for the creation of this paper.

Here, I define three cut-off threshold values to ensure the read out of data quality,
\begin{itemize}
\item P-value cut-off threshold, $L_p$:\\
	For any cluster of centroid-$c$, $\vec x_{i\in c}$, drop the data if $P(\vec x_i|\vec\mu_c) < L_p$.
\item Percentage cut-off threshold, $L_\%$:\\
	For any cluster of centroid-$c$, $x_{i\in c}$, sort the data points according to the P-value in ascending order,
\begin{equation}
\text{sort}\{ P(\vec x_{i\in c}|\vec\mu_c)\},
\end{equation}
	and drop $L_\%$ of data from the begin of the sorted data list.
\item Separation cut-off threshold, $L_s$:\\
	For any data point, $\vec x_{i}$, calculate first and second maximum value of $P(\vec x_i|\vec\mu_c)$ denoted as $P(\vec x_i|\text{1}^{st})$ and $P(\vec x_i|\text{2}^{nd})$. Finally, drop data $\vec x_i$ if the followig criterion matched, 
\begin{equation}
\frac{P(\vec x_i|\text{1}^{st})}{P(\vec x_i|\text{1}^{st})+ P(\vec x_i|\text{2}^{nd})} < L_s.
\end{equation}
\end{itemize}
In general, $L_p$ itself can ensure the data quality and exclude the false classified data points.
However, the simple definition of $L_p$ may cause the imbalance loading of each cluster due to the variance of shapes of each cluster, $P(\vec x|\vec\mu_c)$.
Therefore, $L_\%$ can better ensure a balanced loading of data points in each cluster.
After all, $L_s$ can ensure the separations of two clusters.

\section{Conclusion}
In this paper, I introduced an efficient multi-dimensional clustering algorithm based on the multivariate Gaussian function.
The run time complexity of this new algorithm is much better then the Gaussian mixture model due to the clusters are calculated with locally weighted data points.
On the other hand, similar to Hdbscan algorithms, it automatically find out the locations of each cluster with a better run time complexity (the run time complexity for Hdbscan is roughly $O(N^2 \log N)$).
It would be worth to mention, since most of the operations of this new algorithm are just vector summations, which means it can be easily accelerated with a multi-thread parallel scheme.

While, one can perform a similar calculation by the mixture of {\bf Dbscan/Hdbscan} and {\bf Gaussian mixture model},
\begin{enumerate}
\item run Dbscan/Hdbscan to find out the optimal cluster number and approximate locations of centroids,
\item run Gaussian Mixture model based on the input of cluster number and locations of centroids.
\end{enumerate}
However, due to the run time complexity of Dbscan/Hdbscan and Gaussian mixture model and the difficulties of parallel the algorithm for minimal-spanning-tree, it is not practicall to take this mixed ``two step" algorithm rather then our new algorithm.
I have tested some randomly generated cluster data with 150,000 features, over all, the run time is less than 20 seconds under a 2.5GHz single threaded CPU.
The code is written in C++ with boost library for the need of R-tree data structure.

\end{document}